\documentclass{tran-l}

 \theoremstyle{definition}
 
 \theoremstyle{remark}
 
 \numberwithin{equation}{subsection}

\usepackage{epsfig}
\usepackage{amssymb}
\usepackage{amsmath}
\begin{document}

\title{Distribution of Primes and of Interval Prime Pairs Based on $\Theta$ Function}

\author{ Yifang Fan, Zhiyu Li}

\address[Yifang Fan]{Center for Scientific Research, Guangzhou
Institute of Physical Education, Guangzhou 510500, P.R. China}
\email{tfyf@gipe.edu.cn}
\address[Zhiyu Li]{College of Foreign Languages, Jinan University, Guangzhou
510632, P.R. China}



\begin{abstract}
$\Theta$ function is defined based upon Kronecher symbol. In light
of the principle of inclusion-exclusion, $\Theta$ function of sine
function is used to denote the distribution of composites and
primes. The structure of Goldbach Conjecture has been analyzed, and
$\Xi$ function is brought forward by the linear diophantine
equation; by relating to $\Theta$ function, the interval
distribution of composite pairs and prime pairs (i.e. the Goldbach
Conjecture) is thus obtained. In the end, Abel's Theorem
(Multiplication of Series) is used to discuss the lower limit of the
distribution of the interval prime pairs.
\end{abstract}

\maketitle

\section{Introduction}
\label{sec1} Distribution of primes has long been one of the main
issues of number theory research. Gauss and Legendre presented the
issue of the distribution of primes in the form of conjecture; in
1837, Dirichlet proved the distribution of primes in Arithmetic
Progressions \cite{1}; in 1896, both Hadamard and Poussin proved the basic
laws of the distribution of primes - the prime number theorem; in
1948, Selberg and Erd\"{o}s both proved the prime number theorem by
elementary proofs \cite{2,3}. But still, the present prime number theorem
cannot denote the exact number of primes of natural number.

In 1742, Goldbach brought forward his famous Goldbach Conjecture. Up
to now, the best developments are: in 1937, Vinogradov proved that
each large odd number was the sum of three odd primes \cite{2,4}; in 1966,
Chen Jingrun proved that each large even number can be denoted as
the sum of one prime number and the product of no more than two
primes to multiply \cite{5,6}. In this paper, issues of the distribution
of primes and of the interval prime pairs are discussed by defining
$\Theta$ function and $\Xi$ function. Hereof, \textit{p,q,r} stand
for primes, and $\lceil x\rceil$ is the greatest integer less than
or equal to $x$. A couple of important theorems referred in this
paper include:

\noindent{\bf Theorem 1.} If $N$ is a composite and $p$ is its
smallest positive divisor, then \cite{7}
\begin{displaymath}
p\leq\sqrt{N}
\end{displaymath}
\noindent{\bf Theorem 2.} (Prime Number Theorem) Let $N$ be
composite and $\pi(N)$ denote the number of primes smaller than, it
follows that \cite{2,7}
\begin{displaymath}
\lim\pi(N)=\frac{N}{\ln N}
\end{displaymath}
\noindent{\bf Theorem 3.} (Principle of Inclusion-Exclusion).
Suppose that the number of elements is $n$, where element
$n_{\alpha}$ has the property of $\alpha$, and element $n_{\beta}$
has that of $\beta$, element $n_{\alpha\beta}$ has the properties of
both $\alpha,\beta\cdots$, element $n_{\alpha\beta\gamma}$ has all
the properties of $\alpha,\beta,\gamma\cdots$ , then the number of
elements possessing none of the $\alpha,\beta,\gamma\cdots$
properties is[2, p9; 7, p525]
\begin{displaymath}
n-n_{\alpha}-n_{\beta}-n_{\gamma}-\cdots+n_{\alpha\beta}+\cdots-n_{\alpha\beta\gamma}-\cdots+\cdots-\cdots
\end{displaymath}

\section{Distribution of primes}
Prime number theorem has presented the basic properties of the
distribution of primes. As for a natural number, the result from the
prime number theorem may not be an exact value, but the error will
be diminished as $N$ becomes a larger number \cite{2}. Some
definitions and theorems will be employed to illustrate this
function.
\subsection{Use $\lceil x\rceil$ to denote the distribution of primes and composites}
\noindent{\bf Theorem 4.} When $N$ is an even composite, and $\pi
(N)$ denotes the number of primes no greater than N, then it shows
that \cite{7,8}
\begin{equation}
\label{eq:1}
\pi(N)=N-1-\sum_{i=1}^{l}\left\lceil\frac{N-p_{i}}{p_{i}}\right\rceil+\sum_{i=1}^{l-1}\sum_{j=i+1}^{l}\left\lceil\frac{N}{p_{i}p_{j}}\right\rceil-\sum_{i=1}^{l-2}\sum_{j=i+1}^{l-1}\sum_{k=j+1}^{l}\left\lceil\frac{N}{p_{i}p_{j}p_{k}}\right\rceil+\cdots-\cdots
\end{equation}
Where $p_{i}\leq\sqrt{N}$, $l=\pi(\sqrt{N})$and
 $p_{i}p_{j}\leq{N}, p_{i}p_{j}p_{k}\leq{N}\cdots$

\textbf{Proof.} By Theorem 3, it shows that
\begin{equation}
\label{eq:2}
\pi(N)=N-1+\pi(\sqrt{N})-\sum_{i=1}^{l}\left\lceil\frac{N}{p_{i}}\right\rceil+\sum_{i=1}^{l-1}\sum_{j=i+1}^{l}\left\lceil\frac{N}{p_{i}p_{j}}\right\rceil-\sum_{i=1}^{l-2}\sum_{j=i+1}^{l-1}\sum_{k=j+1}^{l}\left\lceil\frac{N}{p_{i}p_{j}p_{k}}\right\rceil+\cdots-\cdots\
\end{equation}
By Theorem 1, for $p_{i}\leq\sqrt{N}$, it shows that
\begin{equation}
\label{eq:3}
\sum_{i=1}^{l}\left\lceil\frac{N-p_{i}}{p_{i}}\right\rceil=\sum_{i=1}^{l}\left\lceil\frac{N}{p_{i}}\right\rceil-\pi(\sqrt{N})
\end{equation}
Thus the theorem is proved.

\noindent{\bf Theorem 5.} When $N$ is an even composite, and
$\varpi(N)$ denotes the number of composites no greater than $N$, it
shows that
\begin{equation}
\label{eq:4}
\varpi(N)=\sum_{i=1}^{l}\left\lceil\frac{N-p_{i}}{p_{i}}\right\rceil-\sum_{i=1}^{l-1}\sum_{j=i+1}^{l}\left\lceil\frac{N}{p_{i}p_{j}}\right\rceil+\sum_{i=1}^{l-2}\sum_{j=i+1}^{l-1}\sum_{k=j+1}^{l}\left\lceil\frac{N}{p_{i}p_{j}p_{k}}\right\rceil-\cdots+\cdots
\end{equation}
\textbf{Proof.} By the definition of positive integer, it follows
that
\begin{equation}
\label{eq:5} N=\varpi(N)+\pi(N)+1
\end{equation}
By Theorem 4, the proposition is proved.

\subsection{ $\Theta$ function}
\label{sec4}

\noindent{\bf Definition 1.} $\Theta$ function has the following
properties as long as real numbers are concerned
\begin{equation}
\label{eq:6} \Theta (x)= \left\{ \begin{array}{ll}
1 \hspace{1.0cm}& x=0\\
0 & x\neq0
\end{array} \right.
\end{equation}
Let
\begin{displaymath}
\Theta (x)=\delta_{ix}
\end{displaymath}
Where $\delta_{ix}$ is Kronecher symbol, that is
\begin{displaymath}
\delta_{ix}= \left\{ \begin{array}{ll}
1 \hspace{1.0cm}& i=x\\
0 & i\neq x
\end{array} \right.
\end{displaymath}
Suppose $i=0$, then $\Theta (x)$ can be denoted as
\begin{displaymath}
\Theta (x)=\delta_{0x}
\end{displaymath}
$\Theta$ function shows that when $x=0$, the $\Theta$ functional
value of $\Theta$ is $1$; in other cases, it is $0$. $\Theta$ is a
given Kronecher symbol.
\subsection{Use $\Theta$ function of sine function to denote the distribution of primes and that of composites}
\label{sec5} For an even composite, $\lceil x\rceil$ of Theorem 4
and Theorem 5 can be substituted by $\Theta$ function of Definition
1.

\noindent{\bf Theorem 6.} When $N$ is an even composite, use
$\varpi_p (N)$ to denote the number of composites in $N$ that can be
exactly divided by prime $p$, then
\begin{equation}
\label{eq:7}
\varpi_p(N)=\left\lceil\frac{N-p}{p}\right\rceil=\left\lceil\frac{N}{p}\right\rceil-1
\end{equation}
$\Theta$ function of sine function can be denoted as
\begin{equation}
\label{eq:8} \varpi_p(N)=\sum_{x=1}^{N} \Theta
\left(\sin{\left(\frac{x\pi}{p}\right)}\right)-1
\end{equation}
\textbf{Proof.} By the properties of sine function, it follows that
\begin{equation}
\label{eq:9}\\
\underset{1\leqslant x \leqslant
p}{\sin{\left(\frac{x\pi}{p}\right)}}=\left({y_{1},y_{2},\cdots,y_{p}}\right)
\end{equation}
When $x$ is positive integer, then
\begin{equation}
\label{eq:10} \left\{ \begin{array}{ll}
y_{1}\neq0,y_{2}\neq0,\cdots,y_{p-1}\neq0\\
y_{p}=0
\end{array} \right.
\end{equation}
By Definition 1
\begin{equation}
\label{eq:11} \Theta{\left(\underset{1\leqslant x \leqslant
p}{\sin{\left(\frac{x\pi}{p}\right)}}\right)}=\Theta\left({y_{1},y_{2},\cdots,y_{p}}\right)=\left({0,0,\cdots,1}\right)
\end{equation}
That is
\begin{equation}
\label{eq:12} \sum_{x=1}^{p} \Theta
\left(\sin{\left(\frac{x\pi}{p}\right)}\right)=1
\end{equation}
In the same way, it shows that
\begin{equation}
\label{eq:13} \sum_{x=p+1}^{2p} \Theta
\left(\sin{\left(\frac{x\pi}{p}\right)}\right)=1,\cdots,\sum_{x=(n-1)p+1}^{np}
\Theta \left(\sin{\left(\frac{x\pi}{p}\right)}\right)=1
\end{equation}
Where $n$ is a positive integer and $np\leqslant N$.\\
Therefore
\begin{displaymath}
\sum_{x=1}^{N} \Theta
\left(\sin{\left(\frac{x\pi}{p}\right)}\right)=\left\lceil\frac{N}{p}\right\rceil
\end{displaymath}
and by eq.(7), it follows that

\noindent{\bf Theorem 7.} When $N$ is an even composite, use
$\varpi_{pq} (N)$ to denote the number of composites in $N$ that can
be exactly divided by primes $p,q$, then
\begin{equation}
\label{eq:14}
\varpi_{pq}(N)=\left\lceil\frac{N}{p}\right\rceil+\left\lceil\frac{N}{p}\right\rceil-\left\lceil\frac{N}{pq}\right\rceil-2
\end{equation}
$\Theta$ function of sine function can be denoted as
\begin{equation}
\label{eq:15} \varpi_{pq}(N)=\sum_{x=1}^{N}\Theta
\left(\sin{\left(\frac{x\pi}{p}\right)}\right)+\sum_{x=1}^{N}\Theta
\left(\sin{\left(\frac{x\pi}{q}\right)}\right)-\sum_{x=1}^{N}\Theta
\left(\sin{\left(\frac{x\pi}{pq}\right)}\right)-2
\end{equation}
\textbf{Proof.} By the properties of sine function, it follows that
\begin{equation}
\label{eq:16}\\
\underset{1\leqslant x \leqslant
pq}{\sin{\left(\frac{x\pi}{pq}\right)}}=\left({y_{1},y_{2},\cdots,y_{pq}}\right)
\end{equation}
When $x$ is positive integer, it follows that
\begin{equation}
\label{eq:17} \left\{ \begin{array}{ll}
y_{1}\neq0,y_{2}\neq0,\cdots,y_{pq-1}\neq0\\
y_{pq}=0
\end{array} \right.
\end{equation}
By Definition 1
\begin{equation}
\label{eq:18} \Theta{\left(\underset{1\leqslant x \leqslant
pq}{\sin{\left(\frac{x\pi}{pq}\right)}}\right)}=\Theta\left({y_{1},y_{2},\cdots,y_{pq}}\right)=\left({0,0,\cdots,1}\right)
\end{equation}
That is
\begin{equation}
\label{eq:19} \sum_{x=1}^{pq} \Theta
\left(\sin{\left(\frac{x\pi}{pq}\right)}\right)=1
\end{equation}
In the same way, it shows that
\begin{equation}
\label{eq:20} \sum_{x=p+1}^{2pq} \Theta
\left(\sin{\left(\frac{x\pi}{pq}\right)}\right)=1,\cdots,\sum_{x=(n-1)pq+1}^{npq}
\Theta \left(\sin{\left(\frac{x\pi}{pq}\right)}\right)=1
\end{equation}
Where $n$ is a positive integer and $npq\leqslant N$.\\
Therefore
\begin{equation}
\label{eq:21}
 \sum_{x=1}^{N} \Theta
\left(\sin{\left(\frac{x\pi}{pq}\right)}\right)=\left\lceil\frac{N}{pq}\right\rceil
\end{equation}
By Theorem 6, the theorem is proved.

\noindent{\bf Theorem 8.} When $N$ is an even composite, use
$\varpi_{pq} (N)$ to denote the number of composites in $N$ that can
be exactly divided by primes $p,q$, then $\Theta$ function of sine
function can be denoted as
\begin{equation}
\label{eq:22} \varpi_{pq} (N)=\sum_{x=1}^{N}\Theta
\left(\sin{\left(\frac{x\pi}{p}\right)}\sin{\left(\frac{x\pi}{q}\right)}\right)-2
\end{equation}
\textbf{Proof.} Since
\begin{equation}
\label{eq:23}
\left(\sin{\left(\frac{x\pi}{p}\right)}\right)=0\Longrightarrow\Theta\left(\sin{\left(\frac{x\pi}{p}\right)}\sin{\left(\frac{x\pi}{q}\right)}\right)=1
\end{equation}
\begin{equation}
\label{eq:24}
\left(\sin{\left(\frac{x\pi}{q}\right)}\right)=0\Longrightarrow\Theta\left(\sin{\left(\frac{x\pi}{p}\right)}\sin{\left(\frac{x\pi}{q}\right)}\right)=1
\end{equation}
\begin{equation}
\label{eq:25} \left(\sin{\left(\frac{x\pi}{p}\right)}\right)=0,and
\left(\sin{\left(\frac{x\pi}{q}\right)}\right)=0
\Longrightarrow\Theta\left(\sin{\left(\frac{x\pi}{p}\right)}\sin{\left(\frac{x\pi}{q}\right)}\right)=1
\end{equation}
Therefore, it follows that
\begin{equation}
\label{eq:26}
\begin{split}
\sum_{x=1}^{N}\Theta
\left(\sin{\left(\frac{x\pi}{p}\right)}\sin{\left(\frac{x\pi}{q}\right)}\right)
 & = \sum_{x=1}^{N}\Theta \left(\sin{\left(\frac{x\pi}{p}\right)}\right)
+\sum_{x=1}^{N}\Theta \left(\sin{\left(\frac{x\pi}{q}\right)}\right)\\
&-\sum_{x=1}^{N}\Theta
\left(\sin{\left(\frac{x\pi}{pq}\right)}\right)
\\
&
=\left\lceil\frac{N}{p}\right\rceil+\left\lceil\frac{N}{q}\right\rceil-\left\lceil\frac{N}{pq}\right\rceil
\end{split}
\end{equation}
By Theorem 7, the proposition is proved.

\noindent{\bf Theorem 9.} When N is an even composite,
$p,q,\cdots,s\leqslant\sqrt{N}$, and $p\times q\times\cdots \times s
\leqslant{N}$, then.
\begin{equation}
\label{eq:27} \left\lceil\frac{N}{pq\cdots
s}\right\rceil=\sum_{x=1}^{N}\Theta
\left(\sin{\left(\frac{x\pi}{pq\cdots s}\right)}\right)
\end{equation}
\textbf{Proof.} By the properties of sine function, it follows that
\begin{equation}
\label{eq:28}\\
\underset{1\leqslant x \leqslant pq\cdots
s}{\sin{\left(\frac{x\pi}{pq\cdots
s}\right)}}=\left({y_{1},y_{2},\cdots,y_{pq\cdots s}}\right)
\end{equation}
When $x$ is positive integer, it follows that
\begin{equation}
\label{eq:29} \left\{ \begin{array}{ll}
y_{1}\neq0,y_{2}\neq0,\cdots,y_{pq\cdots s-1}\neq0\\
y_{pq\cdots s}=0
\end{array} \right.
\end{equation}
By Definition 1
\begin{equation}
\label{eq:30} \Theta{\left(\underset{1\leqslant x \leqslant pq\cdots
s}{\sin{\left(\frac{x\pi}{pq\cdots
s}\right)}}\right)}=\Theta\left({y_{1},y_{2},\cdots,y_{pq\cdots
s}}\right)=\left({0,0,\cdots,1}\right)
\end{equation}
That is
\begin{equation}
\label{eq:31} \sum_{x=1}^{pq\cdots s} \Theta
\left(\sin{\left(\frac{x\pi}{pq\cdots s}\right)}\right)=1
\end{equation}
In the same way, it follows that
\begin{equation}
\label{eq:32}
\begin{split}
\sum_{x=pq\cdots s+1}^{2pq\cdots s} \Theta
\left(\sin{\left(\frac{x\pi}{pq\cdots
s}\right)}\right)=1,\sum_{x=2pq\cdots s+1}^{3pq\cdots s}
\Theta\left(\sin{\left(\frac{x\pi}{pq\cdots
s}\right)}\right)=1,\cdots,\\
\sum_{x=(n-1)pq\cdots s+1}^{npq\cdots s} \Theta
\left(\sin{\left(\frac{x\pi}{pq\cdots s}\right)}\right)=1
\end{split}
\end{equation}
Therefore, the proposition is proved.

\noindent{\bf Theorem 10.} When $N$ is an even composite, $\varpi
(N)$ of the number of composites smaller than N can be denoted as
\begin{equation}
\label{eq:33} \varpi
(N)=\underset{p_{i}\leqslant\sqrt{N},l=\pi({\sqrt{N}})}{\sum_{x=1}^N
\Theta
\left(\prod_{i=1}^{l}\sin{\left(\frac{x\pi}{p_{i}}\right)}\right)}-\pi(\sqrt{N})
\end{equation}
\textbf{Proof.} By Theorem 5, together with Theorem 6 and 7, can be
denoted as
\begin{equation}
\label{eq:34} \varpi
(N)=\underset{p_{i}\leqslant\sqrt{N}}{\sum_{i=1}^{l}\sum_{x=1}^N
\Theta
\left(\sin{\left(\frac{x\pi}{p_{i}}\right)}\right)}-\underset{p_{i}\leqslant\sqrt{N},p_{i}p_{j}\leqslant{N}}{\sum_{i=1}^{l-1}\sum_{j=i+1}^{l}\sum_{x=1}^N
\Theta
\left(\sin{\left(\frac{x\pi}{p_{i}p_{j}}\right)}\right)}+\cdots-\pi(\sqrt{N})
\end{equation}
Since
\begin{equation}
\label{eq:35} \underset{p_{i}\leqslant\sqrt{N}}{\prod_{i=1}^{l}
\left(\sin{\left(\frac{x\pi}{p_{i}}\right)}\right)}=\underset{l=\pi(\sqrt{N})}{\left(\sin{\left(\frac{x\pi}{p_{1}}\right)}\sin{\left(\frac{x\pi}{p_{2}}\right)}\cdots\sin{\left(\frac{x\pi}{p_{l}}\right)}\right)}
\end{equation}
by Theorem 8 and 9, it follows that
\begin{equation}
\label{eq:36} \underset{p_{i}\leqslant\sqrt{N}}{\sum_{i=1}^N \Theta
\left(\prod_{i=1}^{l}\sin{\left(\frac{x\pi}{p_{i}}\right)}\right)}=\underset{p_{i}\leqslant\sqrt{N}}{\sum_{i=1}^{l}\sum_{x=1}^N
\Theta
\left(\sin{\left(\frac{x\pi}{p_{i}}\right)}\right)}-\underset{p_{i}\leqslant\sqrt{N},p_{i}p_{j}\leqslant{N}}{\sum_{i=1}^{l-1}\sum_{j=i+1}^{l}\sum_{x=1}^N
\Theta
\left(\sin{\left(\frac{x\pi}{p_{i}p_{j}}\right)}\right)}+\cdots
\end{equation}
Therefore, the proposition is proved.

\noindent{\bf Theorem 11.} When $N$ is an even composite, use $\pi
(N)$ to denote the number of primes smaller than $N$, then
\begin{equation}
\label{eq:37} \pi
(N)=N-1-\underset{p_{i}\leqslant\sqrt{N},l=\pi({\sqrt{N}})}{\sum_{x=1}^N
\Theta
\left(\prod_{i=1}^{l}\sin{\left(\frac{x\pi}{p_{i}}\right)}\right)}+\pi(\sqrt{N})
\end{equation}
\textbf{Proof.} When the distribution of composites is determined,
that of the primes can thus be obtained. By Theorem 4 and Theorem
10, prime distribution function is obtained.


\section{Distribution of interval prime pairs}
\subsection{$\Xi$ function}
Take the integer $N$ as an example. We can interpret the Goldbach
Conjecture as follows:
\begin{equation}
\label{eq:38} N=x+y
\end{equation}
The so-called Goldbach Conjecture is that the two numbers that make
up the integer N are both primes. Such primes are called "prime
pair". Now let's denote the continuous positive integers of
$1,2,3,\cdots,N-3,N-2,N-1$ and $N-1,N-2,N-3,\cdots,3,2,1$
respectively with functions. Let's use a function to analyze the
solution of prime pairs of the linear diophantine equation.

\noindent{\bf Definition 2.} $\overset{\rightharpoonup} {\Gamma}$
function, $\overset{\leftharpoonup} {\Gamma}$ function and $\Xi$
function

Suppose that $\overset{\rightharpoonup} {\Gamma}(z)$ is the ordered
function of continuous positive integer of $(1,N)$ and
$\overset{\leftharpoonup} {\Gamma}(z)$ is the ordered function of
continuous positive integer of $(N,1)$, then
$\overset{\rightharpoonup} {\Gamma}(z)+\overset{\leftharpoonup}
{\Gamma}(z)$ will be named as $\Xi$ function of the ordered series
of $\overset{\rightharpoonup} {\Gamma}(z)$ and
$\overset{\leftharpoonup} {\Gamma}(z)$ ,$\overset{\rightharpoonup}
{\Gamma}(z)$ and $\overset{\leftharpoonup} {\Gamma}(z)$ would be an
object to each other. Use $\Xi(z)$ to denote the solution set of two
ordered functions, with possessing the following basic properties:
\begin{enumerate}
 \item Abort $\overset{\rightharpoonup} {\Gamma}(z)$ and $\overset{\leftharpoonup}
 {\Gamma}(z)$, we have
{\begin{equation} \label{eq:39}
\overset{\rightharpoonup}{\Gamma}(z)=x,
\overset{\leftharpoonup}{\Gamma}(z)=N-x=y
\end{equation}}
 \item The solution set of the ordered function of continuous positive integers of (1,N) is
\begin{equation}
\label{eq:40}\Xi(z)=\overset{\rightharpoonup}
{\Gamma}(z)+\overset{\leftharpoonup}{\Gamma}(z)=N
\end{equation}
 \item {Of the results making up $\Xi$ function, two objects will only come up with the following pairings:
1 and composite/prime, composite and composite, composite and prime,
prime and composite, prime and prime, composite/prime and 1. Let's
call the $\Xi$ result making up of prime and prime as the prime pair
(or Goldbach Conjecture), and the rest of them as composite pairs.}
\end{enumerate}

\noindent{\bf Theorem 12.} $\overset
{\rightharpoonup}{\varpi}_{p}(N)$ denotes the number of composites
of $\overset{\rightharpoonup} {\Gamma}(z)$ that can be exactly
divided by $p$ and $\overset {\leftharpoonup}{\varpi}_{p}(N)$
denotes the number of
 composites of $\overset{\leftharpoonup} {\Gamma}(z)$ that can
be exactly divided by $p$. Then for $\overset
{\rightharpoonup}{\varpi}_{p}(N)$ and $\overset
{\leftharpoonup}{\varpi}_{p}(N)$, it follows that
\begin {equation}
\label{eq:41} \overset {\rightharpoonup}{\varpi}_{p}(N)=\overset
{\leftharpoonup}{\varpi}_{p}(N)
\end{equation}
\textbf{Proof. }By Theorem 6, for $\overset
{\rightharpoonup}{\varpi}_{p}(N)$ , it follows that
\begin{equation}
\label{eq:42} \overset {\rightharpoonup}{\varpi}_{p}(N)=\sum_{x=1}^N
\Theta \left(\sin{\left(\frac{x\pi}{p}\right)}\right)-1
\end{equation}
For $\overset {\leftharpoonup}{\varpi}_{p}(N)$, it follows that
\begin{equation}
\label{eq:43} \overset {\leftharpoonup}{\varpi}_{p}(N)=\sum_{x=N}^1
\Theta \left(\sin{\left(\frac{x\pi}{p}\right)}\right)-1
\end{equation}
The theorem is thus proved.

\noindent{\bf Theorem 13.} For the objects of $\overset
{\rightharpoonup}{\Gamma}(N)$ and $\overset
{\leftharpoonup}{\Gamma}(N)$, it follows that
\begin{equation}
\label{eq:44} \overset {\rightharpoonup}{\varpi}(N)=\overset
{\leftharpoonup}{\varpi}(N)
\end{equation}
\begin{equation}
\label{eq:45} \overset {\rightharpoonup}{\pi}(N)=\overset
{\leftharpoonup}{\pi}(N)
\end{equation}
\textbf{Proof. }By Theorem 10, it follows that
\begin{equation}
\label{eq:46} \overset
{\rightharpoonup}{\varpi}(N)=\underset{p_{i}\leqslant\sqrt{N},l=\pi(\sqrt{N})}{\sum_{x=1}^N
\Theta
\left(\sin{\left(\frac{x\pi}{p_{1}}\right)}\sin{\left(\frac{x\pi}{p_{2}}\right)}\cdots\sin{\left(\frac{x\pi}{p_{l}}\right)}\right)}-\pi(\sqrt{N})
\end{equation}
\begin{equation}
\label{eq:47} \overset
{\leftharpoonup}{\varpi}(N)=\underset{p_{i}\leqslant\sqrt{N},l=\pi(\sqrt{N})}{\sum_{x=N}^1
\Theta
\left(\sin{\left(\frac{x\pi}{p_{1}}\right)}\sin{\left(\frac{x\pi}{p_{2}}\right)}\cdots\sin{\left(\frac{x\pi}{p_{l}}\right)}\right)}-\pi(\sqrt{N})
\end{equation}
By Theorem 12, the theorem is proved.
\subsection{Characteristics of $\Xi$ function}
The relationship of $\overset{\rightharpoonup} {\Gamma}(z)$,
$\overset{\leftharpoonup} {\Gamma}(z)$ and $\Xi(z)$ can be
illustrated by the following figure (\ref{fig:1}):

\begin{figure}[!htbp]
 \includegraphics[width=10.5cm]{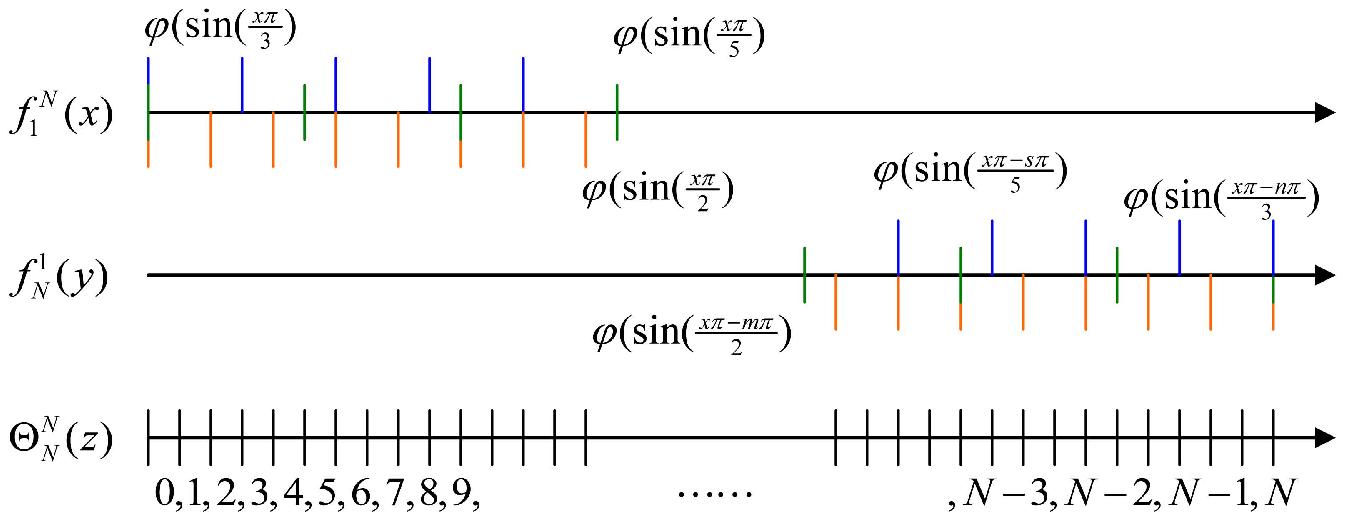}
 \caption{}\label{fig:1}
\end{figure}
By Figure (\ref{fig:1}), we can obtain some characteristics of $\Xi$ function:

\noindent{\bf Theorem 14.} If $N\equiv m(\textrm{mod} p)$, for
$0<m<p$, it follows that
\begin{equation}
\label{eq:48} \overset {\rightharpoonup}{\varpi}_{p}(N)=\underset
{p\leqslant\sqrt{N}}{\sum_{x=1}^N \Theta
\left(\sin{\left(\frac{x\pi}{p}\right)}\right)}=\overset
{\leftharpoonup}{\varpi}_{p}(N)=\underset
{p\leqslant\sqrt{N},N\equiv m(\textrm{mod} p)}{\sum_{x=1}^N \Theta
\left(\sin{\left(\frac{x\pi-m\pi}{p}\right)}\right)}
\end{equation}
\textbf{ Proof.} When $0<m<p$, by function defined by $\Xi(x)$, for
$\overset{\rightharpoonup} {\Gamma}(x)$ and
$\overset{\leftharpoonup} {\Gamma}(x)$, it follows that
\begin{equation}
\label{eq:49} \sum_{x=1}^N \Theta
\left(\sin{\left(\frac{x\pi}{p}\right)}\right)=\sum_{x=1}^N \Theta
\left(\sin{\left(\frac{x\pi-\pi}{p}\right)}\right)=\cdots=\sum_{x=1}^N
\Theta \left(\sin{\left(\frac{x\pi-m\pi}{p}\right)}\right)
\end{equation}
In addition
\begin{equation}
\label{eq:50} \sum_{x=N}^1 \Theta
\left(\sin{\left(\frac{x\pi}{p}\right)}\right)=\sum_{x=1}^N \Theta
\left(\sin{\left(\frac{N\pi-x\pi}{p}\right)}\right)=\sum_{x=1}^N
\Theta \left(\sin{\left(\frac{x\pi-m\pi}{p}\right)}\right)
\end{equation}
Therefore
\begin{equation}
\label{eq:51} \underset {p\leqslant\sqrt{N}}{\sum_{x=1}^N \Theta
\left(\sin{\left(\frac{x\pi}{p}\right)}\right)} =\underset
{p\leqslant\sqrt{N},N\equiv m(\textrm{mod} p)}{\sum_{x=1}^N \Theta
\left(\sin{\left(\frac{x\pi-m\pi}{p}\right)}\right)}
\end{equation}
Together with Theorem 13, the theorem is thus proved.

\noindent{\bf Theorem 15.} When $n,m$ are integers, it follows that
\begin{equation}
\label{eq:52} \Theta
\left(m\sin^{n}{\left(\frac{x\pi}{p}\right)}\right) =\Theta
\left(\sin{\left(\frac{x\pi}{p}\right)}\right)
\end{equation}
\textbf{Proof.} When
\begin{displaymath}
\left(\sin{\left(\frac{x\pi}{p}\right)}\right)\neq0
\end{displaymath}
it follows that
\begin{displaymath}
\left(m\sin^{n}{\left(\frac{x\pi}{p}\right)}\right)\neq0
\end{displaymath}
When
\begin{displaymath}
\left(\sin{\left(\frac{x\pi}{p}\right)}\right)=0
\end{displaymath}
it shows that
\begin{displaymath}
\left(m\sin^{n}{\left(\frac{x\pi}{p}\right)}\right)=0
\end{displaymath}
Hence the proposition holds true.
\subsection{Distribution of composite pairs of $\Xi$ function}
\noindent{\bf Theorem 16.} Let
\begin{displaymath}
x=\Theta \left(\sin{\left(\frac{z\pi}{p}\right)}\right),y=\Theta
\left(\sin{\left(\frac{z\pi}{q}\right)}\right)
\end{displaymath}
for $\Theta$ function, it follows that
\begin{equation}
\label{eq:53} \Theta \left(xy\right) =\Theta \left(x\right)+\Theta
\left(y\right)-\Theta \left(x+y\right)
\end{equation}
\textbf{Proof.} When $p,q$ are primes, the right side shows that
\begin{equation}
\label{eq:54} \Theta (xy)= \left\{ \begin{array}{ll}
1 \hspace{1.0cm}& xy=0\\
0 & xy\neq0
\end{array} \right.
\end{equation}
While the left side shows that
\begin{equation}
\label{eq:55} \left\{ \begin{array}{ll}
xy\neq0\Rightarrow x\neq0,y\neq0\Rightarrow \Theta(x)=0,\Theta(y)=0,\Theta(x+y)=0\\
xy=0 {\left\{ \begin{array}{ll}
x=0,y\neq0\Rightarrow \Theta(x)=1,\Theta(y)=0,\Theta(x+y)=0\\
x\neq0,y=0\Rightarrow \Theta(x)=0,\Theta(y)=1,\Theta(x+y)=0\\
x=0,y=0\Rightarrow \Theta(x)=1,\Theta(y)=1,\Theta(x+y)=1
\end{array} \right.}
\end{array} \right.
\end{equation}
Both sides are equal. Hence the theorem is proved.

\noindent{\bf Theorem 17.} For $\Theta$ function, it follows that
\begin{equation}
\label{eq:56} \underset{p\leqslant \sqrt{N}; N\equiv m(\textrm{mod}
p)}{\Theta\left(\sin{\left(\frac{x\pi}{p}\right)}\sin{\left(\frac{x\pi-m\pi}{p}\right)}\right)}=
{\left\{\begin{array}{ll} \underset{p\leqslant \sqrt{N}; p|N}{\Theta\left(\sin{\left(\frac{x\pi}{p}\right)}\right)}\\
2\underset{p\leqslant \sqrt{N}; p\nmid
N}{\Theta\left(\sin{\left(\frac{x\pi}{p}\right)}\right)}
\end{array} \right.}
\end{equation}
\textbf{Proof.} When $m=0$ that is $p|N$, then
\begin{equation}
\label{eq:57} \underset{p\leqslant \sqrt{N};
p|N}{\Theta\left(\sin{\left(\frac{x\pi}{p}\right)}\sin{\left(\frac{x\pi-m\pi}{p}\right)}\right)}=\underset{p\leqslant
\sqrt{N};p|N}{\Theta\left(\sin{\left(\frac{x\pi}{p}\right)}\sin{\left(\frac{x\pi}{p}\right)}\right)}
\end{equation}
By theorem 15, it shows that
\begin{equation}
\label{eq:58} \underset{p\leqslant
\sqrt{N}}{\Theta\left(\sin{\left(\frac{x\pi}{p}\right)}\sin{\left(\frac{x\pi}{p}\right)}\right)}=\underset{p\leqslant
\sqrt{N}}{\Theta\left(\sin{\left(\frac{x\pi}{p}\right)}\right)}
\end{equation}
Or by Theorem 16, it follows that
\begin{equation}
\label{eq:59}
\begin{split}
\underset{p\leqslant \sqrt{N};
m=0}{\Theta\left(\sin{\left(\frac{x\pi}{p}\right)}\sin{\left(\frac{x\pi-m\pi}{p}\right)}\right)}
& =\underset{p\leqslant
\sqrt{N}}{\Theta\left(\sin{\left(\frac{x\pi}{p}\right)}\right)}+\underset{p\leqslant
\sqrt{N}}{\Theta\left(\sin{\left(\frac{x\pi}{p}\right)}\right)}\\
& -\underset{p\leqslant
\sqrt{N}}{\Theta\left(\sin{\left(\frac{x\pi}{p}\right)}+\sin{\left(\frac{x\pi}{p}\right)}\right)}
\end{split}
\end{equation}
By Theorem 15, it shows that
\begin{equation}
\label{eq:60} \underset{p\leqslant
\sqrt{N}}{\Theta\left(\sin{\left(\frac{x\pi}{p}\right)}+\sin{\left(\frac{x\pi}{p}\right)}\right)}=\underset{p\leqslant
\sqrt{N}}{\Theta\left(2\sin{\left(\frac{x\pi}{p}\right)}\right)}=\underset{p\leqslant
\sqrt{N}}{\Theta\left(\sin{\left(\frac{x\pi}{p}\right)}\right)}
\end{equation}
When $m\neq0$, that is $p\nmid N$, then by theorem 16
\begin{equation}
\label{eq:61}
\begin{split}
\underset{p\leqslant \sqrt{N};p\nmid N;
m\neq0}{\Theta\left(\sin{\left(\frac{x\pi}{p}\right)}\sin{\left(\frac{x\pi-m\pi}{p}\right)}\right)}
& =\underset{p\leqslant
\sqrt{N}}{\Theta\left(\sin{\left(\frac{x\pi}{p}\right)}\right)}+\underset{p\leqslant
\sqrt{N}}{\Theta\left(\sin{\left(\frac{x\pi-m\pi}{p}\right)}\right)}\\
& -\underset{p\leqslant
\sqrt{N}}{\Theta\left(\sin{\left(\frac{x\pi}{p}\right)}+\sin{\left(\frac{x\pi-m\pi}{p}\right)}\right)}
\end{split}
\end{equation}
Since $m\neq0$, $\sin{\left(\frac{x\pi}{p}\right)}$ and
$\sin{\left(\frac{x\pi-m\pi}{p}\right)}$ will not be zero at the
same time. Then it shows that
\begin{equation}
\label{eq:62}
\sin{\left(\frac{x\pi}{p}\right)}+\sin{\left(\frac{x\pi-m\pi}{p}\right)}\neq0
\end{equation}
That is
\begin{equation}
\label{eq:63} \underset{p\leqslant \sqrt{N};p\nmid N;
m\neq0}{\Theta\left(\sin{\left(\frac{x\pi}{p}\right)}+\sin{\left(\frac{x\pi-m\pi}{p}\right)}\right)}=0
\end{equation}
By Theorem 14, it follows that
\begin{equation}
\label{eq:64}
\begin{split}
\underset{p\leqslant \sqrt{N};p\nmid N;
m\neq0}{\Theta\left(\sin{\left(\frac{x\pi}{p}\right)}\sin{\left(\frac{x\pi-m\pi}{p}\right)}\right)}
& =\underset{p\leqslant
\sqrt{N}}{\Theta\left(\sin{\left(\frac{x\pi}{p}\right)}\right)}+\underset{p\leqslant
\sqrt{N}}{\Theta\left(\sin{\left(\frac{x\pi-m\pi}{p}\right)}\right)}\\
& =2\underset{p\leqslant
\sqrt{N}}{\Theta\left(\sin{\left(\frac{x\pi}{p}\right)}\right)}
\end{split}
\end{equation}
The theorem is thus proved.

 \noindent{\bf Theorem 18.} For $\Theta$ function, it follows that
\begin{displaymath}
\underset{p\leqslant \sqrt{N};N\equiv m(\textrm{mod} p); N\equiv
n(\textrm{mod}q)}{\Theta\left(\sin{\left(\frac{x\pi}{p}\right)}\sin{\left(\frac{x\pi-m\pi}{p}\right)}\sin{\left(\frac{x\pi}{q}\right)}\sin{\left(\frac{x\pi-n\pi}{q}\right)}\right)}
\end{displaymath}
\begin{equation}
\label{eq:65} ={\left\{\begin{array}{ll}
 \underset{p\leqslant \sqrt{N};p|N;q|N}{\Theta\left(\sin{\left(\frac{x\pi}{p}\right)}\sin{\left(\frac{x\pi}{q}\right)}\right)}
\\
\underset{p\leqslant \sqrt{N}; p\nmid N;q\nmid
N}{2\Theta\left(\sin{\left(\frac{x\pi}{p}\right)}\right)+2\Theta\left(\sin{\left(\frac{x\pi}{q}\right)}\right)-4\Theta\left(\sin{\left(\frac{x\pi}{pq}\right)}\right)}
\end{array} \right.}
\end{equation}
\textbf{Proof.} When $p,q|N$, it shows that
\begin{equation}
\begin{split}
\label{eq:66} & \underset{p\leqslant \sqrt{N};m=0;
n=0)}{\Theta\left(\sin{\left(\frac{x\pi}{p}\right)}\sin{\left(\frac{x\pi-m\pi}{p}\right)}\sin{\left(\frac{x\pi}{q}\right)}\sin{\left(\frac{x\pi-n\pi}{q}\right)}\right)}\\
& =\underset{p\leqslant \sqrt{N};m=0;
n=0)}{\Theta\left(\sin^2{\left(\frac{x\pi}{p}\right)}\sin^2{\left(\frac{x\pi}{q}\right)}\right)}
\end{split}
\end{equation}
By Theorem 15, this theorem is proved.

When $p,q\nmid N$, it shows that
\begin{displaymath}
\Theta\left(\sin{\left(\frac{x\pi}{p}\right)}+\sin{\left(\frac{x\pi}{q}\right)}\right)
;\Theta\left(\sin{\left(\frac{x\pi-m\pi}{p}\right)}+\sin{\left(\frac{x\pi-n\pi}{q}\right)}\right);
\end{displaymath}
\begin{equation}
\label{eq:67}
\Theta\left(\sin{\left(\frac{x\pi}{p}\right)}+\sin{\left(\frac{x\pi-n\pi}{q}\right)}\right)
;\Theta\left(\sin{\left(\frac{x\pi-m\pi}{p}\right)}+\sin{\left(\frac{x\pi}{q}\right)}\right)
\end{equation}
By Theorem 15 and 16, it shows that
\begin{displaymath}
\underset{p\leqslant \sqrt{N};N\equiv m(\textrm{mod} p); N\equiv
n(\textrm{mod}q)}{\Theta\left(\sin{\left(\frac{x\pi}{p}\right)}\sin{\left(\frac{x\pi-m\pi}{p}\right)}\sin{\left(\frac{x\pi}{q}\right)}\sin{\left(\frac{x\pi-n\pi}{q}\right)}\right)}
\end{displaymath}
\begin{equation}
\label{eq:68} =\underset{p\leqslant \sqrt{N}; p\nmid N;q\nmid
N}{2\Theta\left(\sin{\left(\frac{x\pi}{p}\right)}\right)+2\Theta\left(\sin{\left(\frac{x\pi}{q}\right)}\right)-4\Theta\left(\sin{\left(\frac{x\pi}{pq}\right)}\right)}
\end{equation}
This theorem is thus proved.

\noindent{\bf Theorem 19.} When $p_{i}|N$, use
$\overset{\frown}{\varpi}{(N)}$ to denote the number of composite
pairs of $\Xi$ function. Then
\begin{equation}
\label{eq:69} \underset{p_{i}|N,p_{i}\leqslant\sqrt{N}}
{\overset{\frown}{\varpi}{(N)}}-\pi(\sqrt{N})=\underset{p_{i}|N,p_{i}\leqslant\sqrt{N}}{\overset{\rightharpoonup}{\varpi}{(N)}}=\underset{p_{i}|N,p_{i}\leqslant\sqrt{N}}{\overset{\leftharpoonup}{\varpi}{(N)}}
\end{equation}
That is
\begin{equation}
\label{eq:70} \underset{p_{i}|N,p_{i}\leqslant\sqrt{N}}
{\overset{\frown}{\varpi}{(N)}}=\underset
{p_{i}|N,p_{i}\leqslant\sqrt{N},l=\pi(\sqrt{N})}{\sum_{x=1}^N \Theta
\left(\prod_{i=1}^{l}\sin{\left(\frac{x\pi}{p_{i}}\right)}\right)}
\end{equation}
Wherein the composite pair refers to an order of one or two
composites that correspond to the order of
$\overset{\rightharpoonup}{\Gamma}$ and
$\overset{\leftharpoonup}{\Gamma}$.

\textbf{Proof. }By Theorem 17 and 18, since $p_{i}\mid N$, each
$p_{i}$ of $\overset{\rightharpoonup}{\varpi}{(N)}$ and
$\overset{\leftharpoonup}{\varpi}{(N)}$ will be repeated, then
$p_{i}\leq \sqrt{N}$ will be repeated by
$\underset{p_{i}|N}{\overset{\rightharpoonup}{\varpi}{(N)}}$; in the
same way, $p_{i}\leq \sqrt{N}$ will be repeated by
$\underset{p_{i}|N}{\overset{\leftharpoonup}{\varpi}{(N)}}$. Thus
the theorem is proved.

\noindent{\bf Theorem 20.} When $p_{i}\nmid N$, use
$\overset{\sim}{\varpi}{(N)}$ to denote the number of composite
pairs of $\Xi(N)$ function. Then
\begin{equation}
\label{eq:71} \overset{\sim}{\varpi}{(N)}=\underset
{p_{i}\leqslant\sqrt{N},l=\pi(\sqrt{N})p_{i}\nmid N,N\equiv
m_{i}(\textrm{mod} p_{i})}{\sum_{x=1}^N \Theta
\left(\prod_{i=1}^{l}\sin{\left(\frac{x\pi}{p_{i}}\right)}\sin{\left(\frac{x\pi-m_{i}\pi}{p_{i}}\right)}\right)}
\end{equation}
\textbf{Proof. }By Theorem 17 and 18
\begin{equation}
\label{eq:72}
\begin{split}
\overset{\sim}{\varpi}{(N)} & =\underset
{p_{i}\leqslant\sqrt{N};p_{i}\nmid
N;l=\pi(\sqrt{N})}{2^1\sum_{i=1}^{l}\sum_{x=1}^N \Theta
\left(\sin{\left(\frac{x\pi}{p_{i}}\right)}\right)}-\underset
{p_{i},p_{j}\leqslant\sqrt{N};p_{i},p_{j}\nmid
N;p_{i}p_{j}\leqslant{N};l=\pi(\sqrt{N})}{2^2\sum_{i=1}^{l-1}\sum_{j=i+1}^{l}\sum_{x=1}^N
\Theta
\left(\sin{\left(\frac{x\pi}{p_{i}p_{j}}\right)}\right)}\\
 & +\underset
{p_{i},p_{j},p_{k}\leqslant\sqrt{N};p_{i},p_{j},p_{k}\nmid
N;p_{i}p_{j}p_{k}\leqslant{N};l=\pi(\sqrt{N})}{2^3\sum_{i=1}^{l-2}\sum_{j=i+1}^{l-1}\sum_{k=j+1}^{l}\sum_{x=1}^N
\Theta
\left(\sin{\left(\frac{x\pi}{p_{i}p_{j}p_{k}}\right)}\right)}-\cdots\\
& =\underset {p_{i}\leqslant\sqrt{N};p_{i}\nmid N;N\equiv
m_{i}(\textrm{mod} p_{i});l=\pi(\sqrt{N})}{\sum_{x=1}^N \Theta
\left(\prod_{i=1}^{l}\sin{\left(\frac{x\pi}{p_{i}}\right)}\sin{\left(\frac{x\pi-m_{i}\pi}{p_{i}}\right)}\right)}
\end{split}
\end{equation}
This theorem is thus proved.

 \noindent{\bf Theorem 21.} Use $\overset{\rightleftharpoons}{\varpi}(x)$ to denote the composites of the interval of $(\sqrt{N},N-\sqrt{N})$, it follows that
\begin{equation}
\label{eq:73} \overset{\rightleftharpoons}{\varpi}(x)=\underset
{p_{i}\leqslant\sqrt{N};N\equiv m_{i}(\textrm{mod}
p_{i});l=\pi(\sqrt{N})}{\sum_{x=\sqrt{N}}^{N-\sqrt{N}} \Theta
\left(\prod_{i=1}^{l}\sin{\left(\frac{x\pi}{p_{i}}\right)}\sin{\left(\frac{x\pi-m_{i}\pi}{p_{i}}\right)}\right)}
\end{equation}

\textbf{Proof.} By the definition of composite pair of $\Xi$
function, it shows that
\begin{equation}
\label{eq:74}
\overset{\rightleftharpoons}{\varpi}(x)=\overset{\frown}{\varpi}(x)+\overset{\sim}{\varpi}(x)
\end{equation}
By Theorem 19 and 20, this theorem is proved.

\noindent{\bf Theorem 22.}Use $\overset{\rightleftharpoons}{\pi}(x)$
to denote the number of prime pairs of the interval of
$(\sqrt{N},N-\sqrt{N})$, then
\begin{equation}
\label{eq:75}
\overset{\rightleftharpoons}{\pi}(x)=(N-2\sqrt{N})-\underset
{p_{i}\leqslant\sqrt{N};N\equiv m_{i}(\textrm{mod}
p_{i});l=\pi(\sqrt{N})}{\sum_{x=\sqrt{N}}^{N-\sqrt{N}} \Theta
\left(\prod_{i=1}^{l}\sin{\left(\frac{x\pi}{p_{i}}\right)}\sin{\left(\frac{x\pi-m_{i}\pi}{p_{i}}\right)}\right)}
\end{equation}
\textbf{Proof. } Since $\overset{\rightleftharpoons}{\varpi}(x)$ of
Theorem 21 includes the composite pairs of the interval
$(\sqrt{N},N-\sqrt{N})$, by the definition of $\Xi$ function,we have
\begin{equation}
\label{eq:76}
\overset{\rightleftharpoons}{\varpi}(x)=(N-2\sqrt{N})-\overset{\rightleftharpoons}{\varpi}(x)
\end{equation}
 Hence the theorem is proved.

\section{Further analysis }
\noindent{\bf Theorem 23.} When $N$ is an even composite, use
$\pi(N)$ to denote the number of primes smaller than $N$, then
\begin{equation}
\label{eq:77} \pi(N)=\underset
{p_{i}\leqslant\sqrt{N};l=\pi(\sqrt{N})}{\sum_{x=1}^{N}\Theta\left(\Theta
\left(\prod_{i=1}^{l}\sin{\left(\frac{x\pi}{p_{i}}\right)}\right)\right)}+\pi(\sqrt{N})-1
\end{equation}
\textbf{Proof. }By Theorem 10, it shows that $\underset
{p_{i}\leqslant\sqrt{N};l=\pi(\sqrt{N})}{\sum_{x=1}^{N} \Theta
\left(\prod_{i=1}^{l}\sin{\left(\frac{x\pi}{p_{i}}\right)}\right)}$
 includes the number of composites except $\pi(\sqrt{N})$. When $\underset
{p_{i}\leqslant\sqrt{N};l=\pi(\sqrt{N})}{\prod_{i=1}^{l}\sin{\left(\frac{x\pi}{p_{i}}\right)}}\neq0$,
by the definition of prime, $x$ must be a prime. Then
\begin{equation}
\label{eq:78} \underset
{p_{i}\leqslant\sqrt{N};x\leqslant{N}}{\Theta
\left(\prod_{i=1}^{l}\sin{\left(\frac{x\pi}{p_{i}}\right)}\right)}=0
\end{equation}
Again, by Definition 1, it follows that
\begin{equation}
\label{eq:79} \underset
{p_{i}\leqslant\sqrt{N};l=\pi(\sqrt{N});x\leqslant{N}}{\Theta\left(\Theta
\left(\prod_{i=1}^{l}\sin{\left(\frac{x\pi}{p_{i}}\right)}\right)\right)}=1
\end{equation}
Thus the theorem is proved.

 \noindent{\bf Theorem 24.} Let
\begin{displaymath}
\underset {x\leqslant{N}}{a_{x}}=\underset
{p_{i}\leqslant\sqrt{N};l=\pi(\sqrt{N})}{
\prod_{i=1}^{l}\sin{\left(\frac{x\pi}{p_{i}}\right)}},\underset
{x\leqslant{N}}{b_{x}}=\underset {p_{i}\leqslant\sqrt{N};N\equiv
m_{i}(\textrm{mod} p_{i});l=\pi(\sqrt{N})}{
\prod_{i=1}^{l}\sin{\left(\frac{x\pi-m_{i}\pi}{p_{i}}\right)}}
\end{displaymath}
 for $\Xi$ function, when $a_{x}\neq0$ and $b_{x}\neq0$ , it shows that
\begin{equation}
\label{eq:80} \Theta \left(\Theta \left(a_{x}b_{x}\right)\right)
=\Theta \left(\Theta \left(a_{x}\right)\right)\Theta \left(\Theta
\left(b_{x}\right)\right)
\end{equation}
\textbf{Proof. } When $a_{x}\neq0$ and $b_{x}\neq0$ , then
$a_{x}b_{x}\neq0$. It follows that
\begin{equation}
\label{eq:81} \Theta \left(a_{x}b_{x}\right)=0\Longrightarrow \Theta
\left(\Theta \left(a_{x}b_{x}\right)\right)=1
\end{equation}
\begin{equation}
\label{eq:82} a_{x}b_{x}\neq0\Rightarrow
a_{x}\neq0,b_{x}\neq0\Rightarrow
\begin{array}{ll}
\Theta \left(a_{x}\right)=0,\Theta \left(a_{x}\right)=0\\
\Theta \left(\Theta \left(a_{x}\right)\right)=1;\Theta \left(\Theta\left(b_{x}\right)\right)=1\\
\Theta \left(\Theta\left(a_{x}\right)\right)\Theta \left(\Theta
\left(b_{x}\right)\right)=1\end{array}
\end{equation}
The theorem is thus proved.

\noindent{\bf Theorem 25.} For an even composite N, the number of
prime pairs (Goldbach Conjecture) of interval
$(\sqrt{N},N-\sqrt{N})$ is
\begin{equation}\label{eq:83}
\overset{\rightleftharpoons}{\pi}(x)=\sum_{x=\sqrt{N}}^{N-\sqrt{N}}\underset
{p_{i}\leqslant\sqrt{N};N\equiv m_{i}(\textrm{mod}
p_{i});l=\pi(\sqrt{N}}{ \Theta \left(\Theta
\left(\prod_{i=1}^{l}\sin{\left(\frac{x\pi}{p_{i}}\right)}\right)\right)\Theta
\left(\Theta
\left(\prod_{i=1}^{l}\sin{\left(\frac{x\pi-m_{i}\pi}{p_{i}}\right)}\right)\right)}
\end{equation}
\textbf{Proof.} By the same way of theorem 23, theorem 22 shows that
\begin{equation}\label{eq:84}
\begin{split}
\overset{\rightleftharpoons}{\pi}(x)&
=\sum_{x=\sqrt{N}}^{N-\sqrt{N}}\underset
{p_{i}\leqslant\sqrt{N};N\equiv m_{i}(\textrm{mod}
p_{i});l=\pi(\sqrt{N})}{ \Theta \left(\Theta
\left(\prod_{i=1}^{l}\sin{\left(\frac{x\pi}{p_{i}}\right)\sin{\left(\frac{x\pi-m_{i}\pi}{p_{i}}\right)}}\right)\right)}\\
&=\sum_{x=\sqrt{N}}^{N-\sqrt{N}}\underset
{p_{i}\leqslant\sqrt{N};N\equiv m_{i}(\textrm{mod}
p_{i});l=\pi(\sqrt{N})}{ \Theta \left(\Theta
\left(\prod_{i=1}^{l}\sin{\left(\frac{x\pi}{p_{i}}\right)\prod_{i=1}^{l}\sin{\left(\frac{x\pi-m_{i}\pi}{p_{i}}\right)}}\right)\right)}
\end{split}
\end{equation}
Let
\begin{equation}
\label{eq:85} \underset {p_{i}\leqslant\sqrt{N};l=\pi(\sqrt{N})}{
\prod_{i=1}^{l}\sin{\left(\frac{x\pi}{p_{i}}\right)}}=a_{x},\underset
{p_{i}\leqslant\sqrt{N};N\equiv m_{i}(\textrm{mod}
p_{i});l=\pi(\sqrt{N})}{
\prod_{i=1}^{l}\sin{\left(\frac{x\pi-m_{i}\pi}{p_{i}}\right)}}=b_{x}
\end{equation}
By Theorem 24, theorem 25 is proved.

 \noindent{\bf Theorem 26.} For a sufficiently large even composite of N, it shows that
\begin{equation}\label{eq:86}
\overset{\rightleftharpoons}{\pi}(N)>\frac{N-4\sqrt{N}}{\ln^{2}(N-\sqrt{N})}
\end{equation}
\textbf{Proof.} Since
\begin{equation}\label{eq:87}
\begin{split}
\overset{\rightleftharpoons}{\pi}(x)&=\sum_{x=\sqrt{N}}^{N-\sqrt{N}}\underset
{p_{i}\leqslant\sqrt{N});l=\pi(\sqrt{N})}{ \Theta \left(\Theta
\left(\prod_{i=1}^{l}\sin{\left(\frac{x\pi}{p_{i}}\right)}\right)\right)}
\underset {p_{i}\leqslant\sqrt{N};N\equiv m_{i}(\textrm{mod}
p_{i});l=\pi(\sqrt{N})}{\Theta\left(\Theta\left(\prod_{i=1}^{l}\sin{\left(\frac{x\pi-m_{i}\pi}{p_{i}}\right)}\right)\right)}\\
&=\sum_{x=\sqrt{N}}^{N-\sqrt{N}}\underset
{p_{i}\leqslant\sqrt{N});l=\pi(\sqrt{N})}{ \Theta \left(\Theta
\left(\prod_{i=1}^{l}\sin{\left(\frac{x\pi}{p_{i}}\right)}\right)\right)}
\underset {2<p_{i}\leqslant\sqrt{N};N\equiv m_{i}(\textrm{mod}
p_{i});l=\pi(\sqrt{N})}{\Theta\left(\Theta\left(\prod_{i=1}^{l}\sin{\left(\frac{x\pi-m_{i}\pi}{p_{i}}\right)}\right)\right)}\\
&=\sum_{x=\sqrt{N}}^{N-\sqrt{N}}\underset
{p_{i}\leqslant\sqrt{N});l=\pi(\sqrt{N})}{ \Theta \left(\Theta
\left(\prod_{i=1}^{l}\sin{\left(\frac{x\pi}{p_{i}}\right)}\right)\right)}
\underset {p_{i}\leqslant\sqrt{N};p_{i}\nmid N;N\equiv
m_{i}(\textrm{mod}
p_{i});l=\pi(\sqrt{N})}{\Theta\left(\Theta\left(\prod\_{i=1}^{l}sin{\left(\frac{x\pi-m_{i}\pi}{p_{i}}\right)}\right)\right)}
\end{split}
\end{equation}
Let
\begin{equation}\label{eq:88}
a_{x}=\frac{\underset
{p_{i}\leqslant\sqrt{N};l=\pi(\sqrt{N})}{\Theta\left(\Theta\left(\prod_{i=1}^{l}\sin{\left(\frac{x\pi}{p_{i}}\right)}\right)\right)}}{N-2\sqrt{N}}
\end{equation}
\begin{equation}\label{eq:89}
b_{x}=\frac{\underset {p_{i}\leqslant\sqrt{N};N\equiv
m_{i}(\textrm{mod}
p_{i});l=\pi(\sqrt{N})}{\Theta\left(\Theta\left(\prod_{i=1}^{l}\sin{\left(\frac{x\pi-m_{i}\pi}{p_{i}}\right)}\right)\right)}}{N-2\sqrt{N}}
\end{equation}
\begin{equation}\label{eq:90}
c_{x}=\frac{\underset
{p_{i}\leqslant\sqrt{N};l=\pi(\sqrt{N})}{\Theta\left(\Theta\left(\prod_{i=1}^{l}\sin{\left(\frac{x\pi}{p_{i}}\right)}\right)\right)}
\underset {p_{i}\leqslant\sqrt{N};N\equiv m_{i}(\textrm{mod}
p_{i});l=\pi(\sqrt{N})}{\Theta\left(\Theta\left(\prod_{i=1}^{l}\sin{\left(\frac{x\pi-m_{i}\pi}{p_{i}}\right)}\right)\right)}}{N-2\sqrt{N}}
\end{equation}
By Abel's Theorem (Multiplication of Series) \cite{9}, when
$n\rightarrow \infty$ it follows that
\begin{equation}\label{eq:91}
\begin{split}
&\overset{\rightleftharpoons}{\pi}(x)=(N-2\sqrt{N})\sum_{x=\sqrt{N}}^{N-\sqrt{N}}\frac{\underset
{p_{i}\leqslant\sqrt{N};l=\pi(\sqrt{N})}{\Theta\left(\Theta\left(\prod_{i=1}^{l}\sin{\left(\frac{x\pi}{p_{i}}\right)}\right)\right)}
\underset {p_{i}\leqslant\sqrt{N};N\equiv m_{i}(\textrm{mod}
p_{i});l=\pi(\sqrt{N})}{\Theta\left(\Theta\left(\prod_{i=1}^{l}\sin{\left(\frac{x\pi-m_{i}\pi}{p_{i}}\right)}\right)\right)}}{N-2\sqrt{N}}\\
&=(N-2\sqrt{N})\sum_{x=\sqrt{N}}^{N-\sqrt{N}}\frac{\underset
{p_{i}\leqslant\sqrt{N});l=\pi(\sqrt{N})}{ \Theta \left(\Theta
\left(\prod_{i=1}^{l}\sin{\left(\frac{x\pi}{p_{i}}\right)}\right)\right)}
}{N-2\sqrt{N}} \sum_{x=\sqrt{N}}^{N-\sqrt{N}}\frac{\underset
{p_{i}\leqslant\sqrt{N};N\equiv m_{i}(\textrm{mod}
p_{i})}{\Theta\left(\Theta\left(\prod_{i=1}^{l}\sin{\left(\frac{x\pi-m_{i}\pi}{p_{i}}\right)}\right)\right)}}{N-2\sqrt{N}}\\
&=\frac{\sum_{x=\sqrt{N}}^{N-\sqrt{N}}\underset
{p_{i}\leqslant\sqrt{N});l=\pi(\sqrt{N})}{ \Theta \left(\Theta
\left(\prod_{i=1}^{l}\sin{\left(\frac{x\pi}{p_{i}}\right)}\right)\right)}
\sum_{x=\sqrt{N}}^{N-\sqrt{N}}\underset
{p_{i}\leqslant\sqrt{N};N\equiv m_{i}(\textrm{mod}
p_{i});l=\pi(\sqrt{N})}{\Theta\left(\Theta\left(\prod_{i=1}^{l}\sin{\left(\frac{x\pi-m_{i}\pi}{p_{i}}\right)}\right)\right)}}{(N-2\sqrt{N})}
\end{split}
\end{equation}
Since $N$ is an even number, it surely has $2|N$.

Therefore
\begin{equation}\label{eq:93}
\begin{split}
\sum_{x=\sqrt{N}}^{N-\sqrt{N}}\underset
{p_{i}\leqslant\sqrt{N};p_{i}\nmid{N};N\equiv m_{i}(\textrm{mod}
p_{i})}{\Theta\left(\Theta\left(\prod_{i=1}^{l}\sin{\left(\frac{x\pi-m_{i}\pi}{p_{i}}\right)}\right)\right)}
& \geqslant \sum_{x=\sqrt{N}}^{N-\sqrt{N}}\underset
{2<p_{i}\leqslant\sqrt{N};N\equiv m_{i}(\textrm{mod}
p_{i})}{\Theta\left(\Theta\left(\prod_{i=1}^{l}\sin{\left(\frac{x\pi-m_{i}\pi}{p_{i}}\right)}\right)\right)}\\
& >\sum_{x=\sqrt{N}}^{N-\sqrt{N}}\underset
{p_{i}\leqslant\sqrt{N};N\equiv m_{i}(\textrm{mod}
p_{i})}{\Theta\left(\Theta\left(\prod_{i=1}^{l}\sin{\left(\frac{x\pi-m_{i}\pi}{p_{i}}\right)}\right)\right)}
\end{split}
\end{equation}
So we have
\begin{equation}\label{eq:94}
\overset{\rightleftharpoons}{\pi}(x)>\frac{\sum_{x=\sqrt{N}}^{N-\sqrt{N}}\underset
{p_{i}\leqslant\sqrt{N});l=\pi(\sqrt{N})}{ \Theta \left(\Theta
\left(\prod_{i=1}^{l}\sin{\left(\frac{x\pi}{p_{i}}\right)}\right)\right)}
\sum_{x=\sqrt{N}}^{N-\sqrt{N}}\underset
{p_{i}\leqslant\sqrt{N};N\equiv m_{i}(\textrm{mod}
p_{i});l=\pi(\sqrt{N})}{\Theta\left(\Theta\left(\prod_{i=1}^{l}\sin{\left(\frac{x\pi-m_{i}\pi}{p_{i}}\right)}\right)\right)}}{(N-2\sqrt{N})}
\end{equation}
And
\begin{equation}\label{eq:95}
\sum_{x=\sqrt{N}}^{N-\sqrt{N}}\underset
{p_{i}\leqslant\sqrt{N}}{\Theta\left(\Theta\left(\prod_{i=1}^{l}\sin{\left(\frac{x\pi}{p_{i}}\right)}\right)\right)}=\sum_{x=\sqrt{N}}^{N-\sqrt{N}}\underset
{p_{i}\leqslant\sqrt{N};N\equiv m_{i}(\textrm{mod}
p_{i})}{\Theta\left(\Theta\left(\prod_{i=1}^{l}\sin{\left(\frac{x\pi-m_{i}\pi}{p_{i}}\right)}\right)\right)}
\end{equation}
So
\begin{equation}\label{eq:96}
\overset{\rightleftharpoons}{\pi}(x)>\frac{1}{(N-2\sqrt{N})}\left(\sum_{x=\sqrt{N}}^{N-\sqrt{N}}\underset
{p\leqslant\sqrt{N});l=\pi(\sqrt{N})}{ \Theta \left(\Theta
\left(\prod_{i=1}^{l}\sin{\left(\frac{x\pi}{p_{i}}\right)}\right)\right)}\right)^2
\end{equation}
When $N$ is sufficiently large, by theorem 2, theorem 23 can be
revised as
\begin{equation}\label{eq:97}
\pi(x)=\sum_{x=2}^{N}\underset
{p_{i}\leqslant\sqrt{N});l=\pi(\sqrt{N})}{ \Theta
\left(\Theta\left(\prod_{i=1}^{l}\sin{\left(\frac{x\pi}{p_{i}}\right)}\right)\right)}+\pi(\sqrt{N})=\frac{N}{\ln
N}
\end{equation}
Therefore
\begin{equation}\label{eq:98}
\sum_{x=2}^{N}\underset {p_{i}\leqslant\sqrt{N});l=\pi(\sqrt{N})}{
\Theta
\left(\Theta\left(\prod_{i=1}^{l}\sin{\left(\frac{x\pi}{p_{i}}\right)}\right)\right)}=\frac{N}{\ln
N}-\frac{\sqrt{N}}{\ln \sqrt{N}}=\frac{N-2\sqrt{N}}{\ln N}
\end{equation}
Therefore, the interval of $(\sqrt{N},N-\sqrt{N})$ shows that
\begin{equation}\label{eq:99}
\pi'(N)=\sum_{x=\sqrt{N}}^{N-\sqrt{N}}\underset
{p_{i}\leqslant\sqrt{N});l=\pi(\sqrt{N})}{ \Theta
\left(\Theta\left(\prod_{i=1}^{l}\sin{\left(\frac{x\pi}{p_{i}}\right)}\right)\right)}+\pi(\sqrt{N})=\frac{N-3\sqrt{N}}{\ln
(N-\sqrt{N})}
\end{equation}
Therefore, when $N$ is sufficiently large, it follows that
\begin{equation}\label{eq:100}
\overset{\rightleftharpoons}{\pi}(x)>\frac{N-4\sqrt{N}+\frac{N}{N-2\sqrt{N}}}{\ln^{2}
(N-\sqrt{N})}>\frac{N-4\sqrt{N}}{\ln^{2} (N-\sqrt{N})}
\end{equation}
\section{Discussion}
\begin{figure}[!htbp]
 \includegraphics[width=8.0cm]{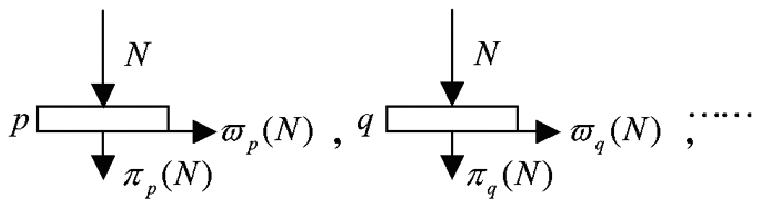}
 \caption{}\label{fig:2}
\end{figure}

Equation (\ref{eq:1}) is a mathematical explanation of sieve method
to Theorem 3. According to Equation (\ref{eq:5}), Equation
(\ref{eq:4}) could mean the same as Equation (\ref{eq:1}). A model
can be applied to explain Equation (\ref{eq:4}). See Figure (\ref{fig:2}).

In Figure (\ref{fig:2}), numbers such as $1, 2,\cdots,N$ will be sifted through
the sieve $p, q, \cdots$ respectively. Here sift is served as a
counting device, not a physical model. (When numbers have been
sifted through the physical model, some numbers will be kept while
others left out.) When numbers are being sifted through sieve p into
q, the numbers remain the same. In order to derive accurate results
from sieves $p, q,$ etc, the issue of repetition between
$\varpi_{p}(N)$ and $\varpi_{q}(N)$ should be taken into account.

\begin{figure}[!htbp]
 \includegraphics[width=5.0cm]{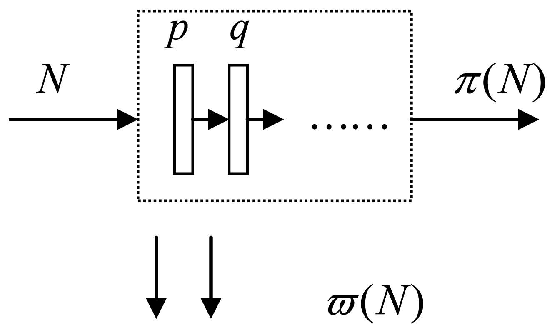}
 \caption{}\label{fig:3}
\end{figure}

 $\Theta$ function can thus be introduced and together with the changing
features of sine function, Equation (\ref{eq:5}) could be rewritten
as Equation (\ref{eq:34}). Comparing Equation (\ref{eq:34}) with
Equation (\ref{eq:5}), we know that Equation (\ref{eq:34}) has
transformed Equation (\ref{eq:5}) into a physical sieve. See Figure
(\ref{fig:3}).

When physical sieve is employed, numbers that could be divided
exactly by p will be left out. The original numbers will be
different from the numbers left after being sifted by p. Thus the
numbers sifted by p will not be repeated by the numbers sifted by q.
Moreover, the number obtained from the remaining ones should be
$\pi(N)$. According to Equation (\ref{eq:5}) and (\ref{eq:34}), we
may be able to get a prime number distribution Equation
(\ref{eq:38}) which is represented by $\Theta$ function of sine
function.

Goldbach Conjecture is a diophantine equation in form. For all
positive whole numbers, when Goldbach Conjecture is presented under
the condition that $N$ is an even number larger than 6, the key to
equation $N=x+y$ would be at least one pair of prime numbers at the
same time, such as $x$ and $y$. The introduction of $\Xi$ function
is to account for a solution to an indefinite equation. We may use
an ordered function to explain the following issue
\begin {displaymath}
(1,N-1)_{1},(1,N-2)_{2},\cdots,(k,N-k)_{k},\cdots,(N-1,1)_{N}
\end{displaymath}
Equations (\ref{eq:44}) and (\ref{eq:45}) tell that "when numbers go
through the physical sieve of Figure (\ref{fig:3}), the result will be related
to the composition of numbers, but irrelated to the order of
numbers, i.e. when numbers of $(1,2,\cdots,N)$ and
$(N,N-1,\cdots,1)$ go through the same physical sieve, the outcome
would remain the same. This can be expressed as
$\overset{\rightharpoonup} {\Gamma}(z)=x$, $\overset{\leftharpoonup}
{\Gamma}(z)=N-x=y$.

Equation (\ref{eq:48}) is one of the key concepts in this paper. By
Figure (\ref{fig:1}), Equation (\ref{eq:48}) can illustrate such an issue that
the sieve to the ordered solution set of $\overset{\rightharpoonup}
{\Gamma}(z)$, and $\overset{\leftharpoonup} {\Gamma}(z)$ of Goldbach
Conjecture will turn out to be the issue of simply sifting the prime
numbers to the ordered solution set of $\overset{\rightharpoonup}
{\Gamma}(z)$, and $\overset{\leftharpoonup} {\Gamma}(z)$.

Equation (\ref{eq:70}) tells us that when $p,q,\cdots$ can divide
$N$ exactly, and if $x$ of the indefinite equation $N=x+y$ can be
divided by $p,q,\cdots$ exactly, then the corresponding $y=N-x$ can
also be divided by $p,q,\cdots$ exactly.

Equation (\ref{eq:73}) has actually constructed such a physical sieve that\\
\begin{figure}[!htbp]
 \includegraphics[width=8.0cm]{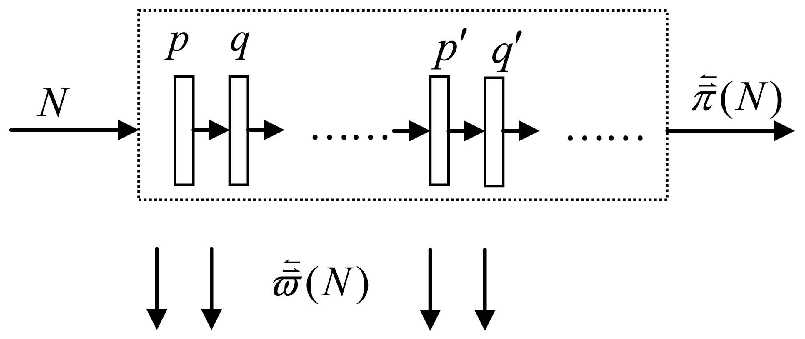}
 \caption{}\label{fig:4}
\end{figure}

The relationship between the sieve of $p,q,\cdots$ and that of
p$p',q',\cdots$ would be that when N can be divided exactly by $p$,
then $p'$ and p are equivalent in figure (\ref{fig:4}). It's just a repetitive sieve which
will not affect the result of sifting. Equation (\ref{eq:52}) serves
as an example of this feature. When $p$ cannot be divided exactly by
N, $p'$ and $p$ are two different siftings. Equations (\ref{eq:73})
and (\ref{eq:34}) have no substantial difference. In the sieve built
by Equation (\ref{eq:73}), the possible prime pairs could be the
ones that have passed the sifting of $p,q,\cdots$(in the ordered
solution set of $\overset{\rightharpoonup} {\Gamma}(x)$, $x$ is a
prime number); the composite number possible prime pairs that cannot
pass the sieve will be sifted again by $p',q',\cdots$(according to
Definition 2, if the solution to an indefinite equation is a
composite number, then it is the solution to the composite number
pair). The pairs that cannot pass the sieve of $p',q',\cdots$ will
be "false" prime pairs (in the ordered collection of
$\overset{\rightharpoonup} {\Gamma}(x)$, $x$ is a prime number while
$y=N-x$ is a composite number. The ones that can pass the sieve of
$p',q',\cdots$ will be real prime pairs (in the ordered solution set
of $\overset{\rightharpoonup} {\Gamma}(x)$, $x$ is a prime number,
and $y=N-x$ is also a prime number).

Goldbach Conjecture can be interpreted as "truth without evidence",
that is to say that under the premise that the conjecture is correct
- the outcome is thus correct. All we have to do is to provide
evidence to prove it.

Equation (\ref{eq:80}) is introduced on the basis of the features of
$\Theta$ function ; with Equation (\ref{eq:80}), Equation
(\ref{eq:75}) is turned into Equation (\ref{eq:83}). With Equation
(\ref{eq:83}), the possibility to prove Equation (\ref{eq:86}) has
become a reality. In Equation (\ref{eq:83}), the multiplication of
progression has come out as a problem. Abel's Theorem is employed to
solve this problem. It's well proved in Equation (\ref{eq:86}).

To sum up, Equation (\ref{eq:86}) can be testified for the following
reasons:

a) It seems to be quite tough to explain the Goldbach Conjecture
only by the conventional sieve since it involved the repetition
issue when counting the numbers of sieves (prime factor). But a
physical sieve has been set up (Equation (\ref{eq:34})) by applying
the sine
function of $\Theta$ Function .

b) By introducing $\Xi$ Function, the Goldbach Conjecture (the
indefinite equation issue) can be simplified as a prime number
sifting issue. Equations (\ref{eq:73}) and (\ref{eq:34}) have no
substantial difference. By applying Equation (\ref{eq:34}), the
composite and prime numbers have been sifted while through Equation
(\ref{eq:73}), the pair numbers of composite and
prime numbers are sifted.

c) With further application of $\Theta$ Function, Equation
(\ref{eq:83}) is thus obtained, which is in line with the
multiplication of progression issue. By applying Abel's Theorem, the
correctness of Goldbach Conjecture is thus proved.

\section*{Acknowledgments}

This project was funded by National Natural Science Foundation of
China under the grant $10772053$, $10972061$ and by Key Project of
Natural Science Research of Guangdong Higher Education Grant No
$06Z019$. The authors would like to acknowledge the support from
the subjects.

\section*{Appendix}

\textbf{Example 1.}(Theorem 4)
\begin {displaymath}
\pi(10)=10-1-\left\lceil\frac{10-2}{2}\right\rceil-\left\lceil\frac{10-3}{3}\right\rceil+\left\lceil\frac{10}{2\times3}\right\rceil=9-4-2+1=4
\end{displaymath}
The fact is that $\pi(10)$ includes 2,3,5,7.

\textbf{Example 2.}(Theorem 5)
\begin{displaymath}
\varpi(10)=\left\lceil\frac{10-2}{2}\right\rceil+\left\lceil\frac{10-3}{3}\right\rceil-\left\lceil\frac{10}{2\times3}\right\rceil=4+2-1=5
\end{displaymath}
The fact is that $\varpi(10)$ includes 4,6,8,9,10.

\textbf{Example3.} (Theorem 6) For integer $10$, by eq.(6), it
follows that
\begin{displaymath}
\varpi_3(10)=\left\lceil\frac{10-3}{3}\right\rceil=\left\lceil\frac{10}{3}\right\rceil-1=2
\end{displaymath}
\begin{displaymath}
\varpi_3(10)=\Theta
\left(\sin{\left(\frac{3\pi}{3}\right)}\right)+\Theta
\left(\sin{\left(\frac{6\pi}{3}\right)}\right)+\Theta
\left(\sin{\left(\frac{9\pi}{3}\right)}\right)-1=3-1=2
\end{displaymath}

\textbf{Example 4.}(Theorem 8)

For integer 10, by formula (3), it follows that
\begin{displaymath}
\varpi_{2,3}(N)=\left\lceil\frac{10}{2}\right\rceil+\left\lceil\frac{10}{3}\right\rceil-\left\lceil\frac{10}{2\times3}\right\rceil-2=5
\end{displaymath}
While by formula (4), it follows that
\begin{displaymath}
\begin{split}
\varpi_{2,3}(N) &
=\Theta\left(\sin{\left(\frac{2\pi}{2}\right)}\right)+\cdots+\Theta\left(\sin{\left(\frac{10\pi}{2}\right)}\right)
\\
&
+\Theta\left(\sin{\left(\frac{3\pi}{3}\right)}\right)+\Theta\left(\sin{\left(\frac{6\pi}{3}\right)}\right)+\Theta\left(\sin{\left(\frac{9\pi}{3}\right)}\right)\\
&
-\Theta\left(\sin{\left(\frac{6\pi}{2\times3}\right)}\right)-2=5+3-1-2=5
\end{split}
\end{displaymath}
And by formula (5), it shows that
\begin{displaymath}
\begin{split}
\varpi_{2,3}(N) &
=\Theta\left(\sin{\left(\frac{2\pi}{2}\right)}\sin{\left(\frac{2\pi}{3}\right)}\right)+\Theta\left(\sin{\left(\frac{3\pi}{2}\right)}\sin{\left(\frac{3\pi}{3}\right)}\right)
\\
&
+\Theta\left(\sin{\left(\frac{4\pi}{2}\right)}\sin{\left(\frac{4\pi}{3}\right)}\right)+\Theta\left(\sin{\left(\frac{6\pi}{2}\right)}\sin{\left(\frac{6\pi}{3}\right)}\right)
\\
&
+\Theta\left(\sin{\left(\frac{8\pi}{2}\right)}\sin{\left(\frac{8\pi}{3}\right)}\right)+\Theta\left(\sin{\left(\frac{9\pi}{2}\right)}\sin{\left(\frac{9\pi}{3}\right)}\right)
\\
&
+\Theta\left(\sin{\left(\frac{10\pi}{2}\right)}\sin{\left(\frac{10\pi}{3}\right)}\right)-2=7-2=5
\end{split}
\end{displaymath}

\textbf{Example 5.}(Theorem 12)
\begin{displaymath}
\begin{split}
\overset {\rightharpoonup}{\varpi}_{3}(10)&=\Theta
\left(\sin{\left(\frac{3\pi}{3}\right)}\right)+\Theta
\left(\sin{\left(\frac{6\pi}{3}\right)}\right)+\Theta
\left(\sin{\left(\frac{9\pi}{3}\right)}\right)-1\\
&=3-1=2
\end{split}
\end{displaymath}

\begin{displaymath}
\begin{split}
\overset {\leftharpoonup}{\varpi}_{3}(10)&=\Theta
\left(\sin{\left(\frac{9\pi}{3}\right)}\right)+\Theta
\left(\sin{\left(\frac{6\pi}{3}\right)}\right)+\Theta
\left(\sin{\left(\frac{3\pi}{3}\right)}\right)-1\\
&=3-1=2
\end{split}
\end{displaymath}

\textbf{Example 6.}(Theorem 17)
\begin{displaymath}
\begin{split}
\sum_{m=1}^{2}\sum_{x=1}^{10}{\Theta\left(\sin{\left(\frac{x\pi-m\pi}{5}\right)}\right)}
&
={\Theta\left(\sin{\left(\frac{4\pi-\pi}{5}\right)}\right)}+{\Theta\left(\sin{\left(\frac{9\pi-\pi}{5}\right)}\right)}\\
&
+{\Theta\left(\sin{\left(\frac{3\pi-2\pi}{5}\right)}\right)}+{\Theta\left(\sin{\left(\frac{8\pi-2\pi}{5}\right)}\right)}=4\\
\end{split}
\end{displaymath}

\begin{displaymath}
2\sum_{x=1}^{10}{\Theta\left(\sin{\left(\frac{x\pi}{5}\right)}\right)}=2\left({\Theta\left(\sin{\left(\frac{5\pi}{5}\right)}\right)+\Theta\left(\sin{\left(\frac{10\pi}{5}\right)}\right)}\right)=4
\end{displaymath}

\textbf{Example 7.}(Theorem 19)
\begin{displaymath}
\begin{split}
\widehat{\varpi}=&{\Theta\left(\sin{\left(\frac{2\pi}{2}\right)}\sin{\left(\frac{2\pi}{3}\right)}\right)}+{\Theta\left(\sin{\left(\frac{3\pi}{2}\right)}\sin{\left(\frac{3\pi}{3}\right)}\right)}+{\Theta\left(\sin{\left(\frac{4\pi}{2}\right)}\sin{\left(\frac{4\pi}{3}\right)}\right)}\\
&
+{\Theta\left(\sin{\left(\frac{6\pi}{2}\right)}\sin{\left(\frac{6\pi}{3}\right)}\right)}+{\Theta\left(\sin{\left(\frac{8\pi}{2}\right)}\sin{\left(\frac{8\pi}{3}\right)}\right)}\\
&
+{\Theta\left(\sin{\left(\frac{9\pi}{2}\right)}\sin{\left(\frac{9\pi}{3}\right)}\right)}+{\Theta\left(\sin{\left(\frac{12\pi}{2}\right)}\sin{\left(\frac{12\pi}{3}\right)}\right)}=7\\
&
\end{split}
\end{displaymath}

\textbf{Example 8. }(Theorem 25 and Theorem 26)

When $N=100$ and $\pi\sqrt{N}=4$, then $100\equiv1(\textrm{mod} 3)$,
$100\equiv2(\textrm{mod} 7)$. The interval of $[10, 90]$ shows that

When
\begin{displaymath}
\overset{\rightharpoonup}{\Gamma}(x)=\sum_{x=10}^{90}\Theta
\left(\Theta
\left(\sin{\left(\frac{x\pi}{2}\right)}\sin{\left(\frac{x\pi}{3}\right)}\sin{\left(\frac{x\pi}{5}\right)}\sin{\left(\frac{x\pi}{7}\right)}\right)\right)=1
\end{displaymath}
$x$ is $11,13,17,19,23,29,31,37,41,43,47,53,59,61,67,71,73,79,83,89$.\\
When
\begin{displaymath}
\overset{\leftharpoonup}{\Gamma}(x)=\sum_{x=10}^{90}\Theta
\left(\Theta
\left(\sin{\left(\frac{x\pi}{2}\right)}\sin{\left(\frac{x\pi-\pi}{3}\right)}\sin{\left(\frac{x\pi}{5}\right)}\sin{\left(\frac{x\pi-2\pi}{7}\right)}\right)\right)=1
\end{displaymath}
$x$ is 11,17,21,27,29,33,39,41,47,53,57,59,63,69,71,77,81,83,87,89.\\

$\Theta \left(\Theta
\left(\sin^2{\left(\frac{x\pi}{2}\right)}\sin{\left(\frac{x\pi}{3}\right)}\sin^2{\left(\frac{x\pi}{5}\right)}\sin{\left(\frac{x\pi}{7}\right)}\sin{\left(\frac{x\pi-\pi}{3}\right)}\sin{\left(\frac{x\pi-2\pi}{7}\right)}\right)\right)$
\\
$=\Theta \left(\Theta
\left(\sin{\left(\frac{x\pi}{2}\right)}\sin{\left(\frac{x\pi}{3}\right)}\sin{\left(\frac{x\pi}{5}\right)}\sin{\left(\frac{x\pi}{7}\right)}\sin{\left(\frac{x\pi-\pi}{3}\right)}\sin{\left(\frac{x\pi-2\pi}{7}\right)}\right)\right)$
\\
When
\begin{displaymath}
\overset{\rightleftharpoons}{\pi}(100)=\sum_{x=10}^{90}\Theta
\left(\Theta\left(\sin{\left(\frac{x\pi}{2}\right)}\cdots\sin{\left(\frac{x\pi-2\pi}{7}\right)}\right)\right)=1
\end{displaymath}
$x$ is $11,17,29,41,47,53,59,71,83,89$.
\begin{displaymath}
\begin{split}
\overset{\rightleftharpoons}{\pi}(100)&=\sum_{x=10}^{90}\Theta
\left(\Theta
\left(\sin{\left(\frac{x\pi}{2}\right)}\sin{\left(\frac{x\pi}{3}\right)}\sin{\left(\frac{x\pi}{5}\right)}\sin{\left(\frac{x\pi}{7}\right)}\right)\right)\\
&\Theta\left(\Theta\left(\sin{\left(\frac{x\pi}{2}\right)}\sin{\left(\frac{x\pi-\pi}{3}\right)}\sin{\left(\frac{x\pi}{5}\right)}\sin{\left(\frac{x\pi-2\pi}{7}\right)}\right)\right)=1
\end{split}
\end{displaymath}
$x$ is $11,17,29,41,47,53,59,71,83,89$.
\begin{displaymath}
\begin{split}
\overset{\rightleftharpoons}{\pi}(100)&=\sum_{x=10}^{90}\Theta
\left(\Theta
\left(\sin{\left(\frac{x\pi}{2}\right)}\sin{\left(\frac{x\pi}{3}\right)}\sin{\left(\frac{x\pi}{5}\right)}\sin{\left(\frac{x\pi}{7}\right)}\right)\right)\\
&\Theta\left(\Theta\left(\sin{\left(\frac{x\pi-\pi}{3}\right)}\sin{\left(\frac{x\pi-2\pi}{7}\right)}\right)\right)=1
\end{split}
\end{displaymath}
$x$ is $11,17,29,41,47,53,59,71,83,89$.

The fact is that $\overset{\rightleftharpoons}{\pi}(100)=10$ and the
specific prime pairs are
\begin{displaymath}
\begin{split}
& 11,17,29,41,47,53,59,71,83,89 \\
& 89,83,71,59,53,47,41,29,17,11
\end{split}
\end{displaymath}
When $N=1000$ and $\pi\sqrt{N}=11$, then $1000\equiv1(\textrm{mod}
3)$, $1000\equiv6(\textrm{mod}7)$,$\cdots$,
$1000\equiv8(\textrm{mod}31)$. The interval of $[32, 968]$ has
$\overset{\rightleftharpoons}{\pi}(1000)=48$, the specific prime
pairs are
\begin{displaymath}
\begin{split}
& 47,53,59,\cdots,443,479,\cdots,941,947,953\\
& 953,947,941,\cdots,557,521,\cdots,59,53, 47
\end{split}
\end{displaymath}
When $N=10000$ and $\pi\sqrt{N}=25$, then $10000\equiv1(\textrm{mod}
3)$,$10000\equiv4(\textrm{mod}7)$,
$\cdots$,$10000\equiv9(\textrm{mod} 97)$. The interval of $[100,
9900]$ has $\overset{\rightleftharpoons}{\pi}(10000)=232$, the
specific prime pairs are
\begin{displaymath}
\begin{split}
&113,149,163,\cdots,4919,5081,\cdots,9837,9851,9887\\
&9887,9851,9837,\cdots,5081,4919,\cdots,163,149, 113
\end{split}
\end {displaymath}


\end{document}